\documentclass[preprint,showpacs,preprintnumbers,
superscriptaddress,prd]{revtex4}

\usepackage{amsmath,amssymb}
\usepackage{graphicx}
\newcommand{\bs}[1]{\ensuremath{\boldsymbol{#1}}}

\begin{document}

\title{Post-Newtonian effects on the stability of the triangular solution 
in the three-body problem for general masses}
\author{Kei Yamada}
\email{k.yamada@tap.scphys.kyoto-u.ac.jp}
\affiliation{
Department of Physics, Kyoto University, Kyoto 606-8502, Japan} 

\author{Takuya Tsuchiya} 
\affiliation{
Department of Mathematical Sciences, Waseda University, 
Shinjuku 169-8555, Japan} 

\author{Hideki Asada} 
\affiliation{
Faculty of Science and Technology, Hirosaki University,
Hirosaki 036-8561, Japan}


\begin{abstract}

Continuing work initiated in earlier publications 
[Ichita, Yamada and Asada, Phys. Rev. D {\bf 83}, 084026 (2011);
Yamada and Asada, Phys. Rev. D {\bf 86}, 124029 (2012)], 
we examine the post-Newtonian (PN) effects on the stability of 
the triangular solution 
in the relativistic three-body problem for general masses.
For three finite masses, 
a condition for stability of the triangular solution is obtained 
at the first post-Newtonian (1PN) order, 
and it recovers 
previous results for the PN restricted three-body problem 
when one mass goes to zero.
The stability regions still exist even at the 1PN order,
though 
the PN triangular configuration for general masses is 
less stable than 
the PN restricted three-body case as well as the Newtonian one.
\end{abstract}


\maketitle

\section{Introduction}

One of classical problems in astronomy and physics is 
the three-body problem in Newtonian gravity 
(e.g., \cite{Goldstein,Danby,Marchal}). 
The gravitational three-body problem is not integrable 
by analytical methods.
As particular solutions, however, 
Euler and Lagrange found a collinear solution and 
an equilateral triangular one, respectively.
The solutions to the restricted three-body problem, 
where one of the three bodies is a test particle, are known as 
Lagrangian points $L_1, L_2, L_3, L_4$ and $L_5$ \cite{Goldstein}. 
Lagrange's equilateral triangular solution has also a practical importance, 
since $L_4$ and $L_5$ for the Sun-Jupiter system are stable 
and indeed the Trojan asteroids are located there. 
Even for the Sun-Earth system, 
asteroids were also found around $L_4$ by recent observations \cite{CWV}.

Recently, Lagrangian points have attracted renewed interests 
for relativistic astrophysics 
\cite{Krefetz, Maindl, SM, Schnittman, Asada, THA}, 
where they have discussed the post-Newtonian (PN) corrections for 
Lagrangian points \cite{Krefetz, Maindl}, and the gravitational radiation 
reaction on $L_4$ and $L_5$ analytically \cite{Asada} 
and by numerical methods \cite{SM,Schnittman,THA}. 
It is currently important to reexamine Lagrangian points in the framework 
of general relativity.
As a pioneering work \cite{Nordtvedt}, 
Nordtvedt has pointed out that 
the location of the triangular points is very sensitive 
to the ratio of the gravitational mass to the inertial one, 
though his analysis does not fully take account of 
the first post-Newtonian (1PN) terms.
Along this course, it might be important as a gravity experiment 
to discuss the three-body coupling terms in the PN force,  
because some of the terms are proportional to a product of 
three masses such as $m_1 \times m_2 \times m_3$. 
Such a triple product appears 
only for relativistic three (or more) body systems 
but it does not for relativistic compact binaries 
nor Newtonian three-body systems. 
In addition, 
it has been pointed out that three-body interactions might play 
important roles for compact binary mergers in hierarchical triple systems 
\cite{BLS, MH, Wen, Thompson, Seto}.
Very recently, moreover, 
a first relativistic hierarchical triple system has been discovered 
by Ransom and his collaborators \cite{Ransom}.

For three finite masses, in the 1PN approximation,
the existence and uniqueness 
of a PN collinear solution corresponding to Euler's one 
have been shown by Yamada and Asada \cite{YA1, YA2}.
Also, 
Ichita et al., including two of the present authors, have shown that 
an equilateral triangular solution is possible at the 1PN order,
if and only if all the three masses are equal \cite{IYA}. 
Generalizing this earlier work, 
Yamada and Asada have found 
a {\it PN triangular} equilibrium solution for general masses
with 1PN corrections to each side length \cite{YA3}.
This PN triangular configuration for general masses is 
not always equilateral
and it recovers the previous results by Krefetz \cite{Krefetz} 
and Maindl \cite{Maindl} 
for the restricted three-body case.

In Newtonian gravity, 
Gascheau proved that 
Lagrange's equilateral triangular configuration for circular motion is stable 
\cite{Gascheau}, if
\begin{align}
\frac{m_1 m_2 + m_2 m_3 + m_3 m_1}{(m_1 + m_2 + m_3)^2} < \frac{1}{27} .
\label{CoSN}
\end{align}
Routh extended the result to a general law of gravitation $\propto 1/r^k$, 
and found the condition for stability as \cite{Routh}
\begin{align}
\frac{m_1 m_2 + m_2 m_3 + m_3 m_1}{(m_1 + m_2 + m_3)^2} < 
\frac13 \left( \frac{3 - k}{1 + k} \right)^2 .
\end{align}
For the restricted three-body limit as $m_3 \to 0$, 
Douskos and Perdios examined the stability of $L_4$ ($L_5$) 
in the 1PN approximation,
and they obtained a region of stability as \cite{DP}
\begin{align}
\frac{m_2}{m_1 + m_2} < \mu_0 - \frac{17 \sqrt{69}}{486} \varepsilon ,
\label{CoS-R3BP}
\end{align} 
where 
we assume $m_1 > m_2$ without loss of generality, 
the Newtonian value $\mu_0 = (9 - \sqrt{69})/18$, 
and we define 
\begin{align}
\varepsilon \equiv \frac{G M}{c^2 \ell} 
\end{align}
for the total mass $M$ and 
each side length $\ell$ of the Newtonian equilateral triangle.
Singh and Bello also discussed the stability 
in the restricted three-body problem 
by taking account of not only the general relativistic effects 
but also radiation pressure and 
a small perturbation in the centrifugal force \cite{SB1,SB2}.
For three finite masses, however, 
it is not clear whether the PN triangular configuration is stable. 
In this paper, 
we study this issue
by {\it fully} taking account of all the 1PN terms. 

This paper is organized as follows. 
In Sec. \ref{Solution}, 
we briefly summarize the PN triangular equilibrium solution 
for three finite masses. 
In Sec. \ref{Stability}, 
we consider the stability of the solution at the 1PN order.
Section \ref{Conclusion} is devoted to the conclusion.
Hereafter, we take the units of $G = c = 1$. 

\section{A Post-Newtonian Triangular Solution}
\label{Solution}

In this section, 
following Ref. \cite{YA3},
we summarize a derivation of the PN triangular solution 
to the relativistic three-body problem for general masses.
We take account of the terms at the 1PN order by employing 
the Einstein-Infeld-Hoffman (EIH) equations of motion 
in the standard PN coordinates as \cite{MTW,LL,Will,AFH} 
\begin{align}
m_K \frac{d^2 \bs{r}_K}{d t^2} 
&= \sum_{A \neq K} \bs{r}_{A K} 
\frac{m_K m_A}{r_{A K}^3} 
\left[ 1 - 4 \sum_{B \neq K} \frac{m_B}{r_{B K}} 
- \sum_{C \neq A} \frac{m_C}{r_{C A}} 
\left( 1 - \frac{\bs{r}_{A K} \cdot \bs{r}_{C A}}{2 r_{C A}^2} \right) 
\right.
\notag\\
&~~~
\left.
+ v_K^2 + 2 v_A^2 - 4 \bs{v}_A \cdot \bs{v}_K 
- \frac32 \left( \bs{v}_A \cdot \bs{n}_{A K} \right)^2 \right]
\notag\\
&~~~
- \sum_{A \neq K} (\bs{v}_A - \bs{v}_K) 
\frac{m_K m_A}{r_{A K}^2} \bs{n}_{A K} \cdot (3 \bs{v}_A - 4 \bs{v}_K)
+ \frac72 \sum_{A \neq K} \sum_{C \neq A} 
\bs{r}_{C A} \frac{m_K m_A m_C}{r_{A K} r_{C A}^3} ,
\label{EIH-EOM}
\end{align}
where $\bs{r}_I$ and $\bs{v}_I$ denote 
the location and the velocity of each body in an inertial frame 
and we define 
\begin{align}
\bs{r}_{I J} &\equiv \bs{r}_I - \bs{r}_J , \\
r_{I J} &\equiv |\bs{r}_{I J}| , \\
\bs{n}_{I J} &\equiv \frac{\bs{r}_{I J}}{r_{I J}} .
\end{align}
In the following, we assume circular motion.

\begin{figure}[ht]
\begin{center}
  \includegraphics[width=10cm]{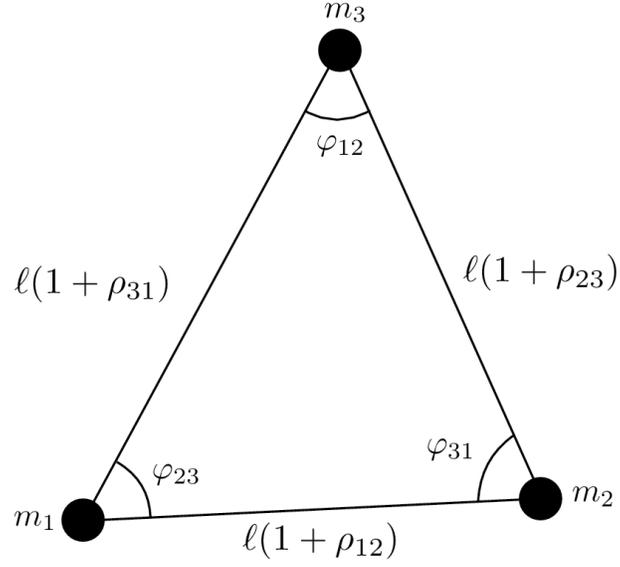}
\caption{PN triangular configuration.
Each body is located at one of the apexes.
$\rho_{I J}$ denotes the PN corrections to each side length at the 1PN order.
In the equilateral case, $\rho_{12} = \rho_{23} = \rho_{31} = 0$, 
namely, $r_{12} = r_{23} = r_{31} = \ell$ 
according to Eq. \eqref{arithmetical-mean}.}
\label{C5fig:fig-PN-tri}
\end{center}
\end{figure}

We consider a PN triangular configuration 
with 1PN corrections to each side length of a Newtonian equilateral triangle, 
so that the distances between the bodies are
\begin{align}
r_{I J} = \ell (1 + \rho_{I J}) ,
\end{align}
where $I, J = 1, 2, 3$ and 
$\rho_{I J} ( = \rho_{J I} )$ is dimensionless PN corrections 
(see Fig. \ref{C5fig:fig-PN-tri}). 
Because of the circular motion, $\ell$ and $\rho_{I J}$ are constants.
Note that we neglect 
the terms of second (and higher) order in $\varepsilon$ henceforth.
Here, if all the three corrections are equal 
(i.e. $\rho_{12} = \rho_{23} = \rho_{31} = \rho$), 
a PN configuration is still an equilateral triangle, 
though each side length is changed by a scale transformation as 
$\ell \to \ell (1 + \rho)$.
Namely, 
one of the degrees of freedom for the PN corrections corresponds to 
a scale transformation, and this is unimportant. 
In order to eliminate this degree of freedom, 
we impose a constraint condition 
\begin{align}
\frac{r_{12} + r_{23} + r_{31}}{3} = \ell ,
\label{arithmetical-mean}
\end{align}
which means that 
the arithmetical mean of the three distances of the bodies is not changed 
by the PN corrections.
Namely, 
\begin{align}
\rho_{12} + \rho_{23} + \rho_{31} = 0 .
\label{cc-1PNcorrection}
\end{align}
Please see also Ref. \cite{YA3} for imposing this constraint.

Let $\omega_I$ denote the angular velocities of the $I$th body 
with PN corrections. 
The EIH equation of motion for $m_1$ becomes
\begin{align}
- \omega_1^2 \boldsymbol{r}_{{\rm N} 1} &= 
- \frac{M}{\ell^3} \boldsymbol{r}_{{\rm N} 1}
+ \boldsymbol{\delta}_{{\rm EIH} 1} \varepsilon 
- \frac{3}{2} \frac{M}{\ell^2} \frac{1}{\sqrt{\nu_2^2 + \nu_2\nu_3 + \nu_3^2}}
\notag\\
&~~~
\times \{ [\nu_2(\nu_1 - \nu_2 - 1)\rho_{12} 
+ \nu_3(\nu_1 - \nu_3 - 1)\rho_{31}]\boldsymbol{n}_{1} 
+ \sqrt{3}\nu_2\nu_3
(\rho_{12} - \rho_{31})\boldsymbol{n}_{\perp 1} \},
\label{C5eq:M1-EOM-C}
\end{align}
where the mass ratio $\nu_I \equiv m_I/M$,
$\bs{r}_{{\rm N} I}$ and $\bs{v}_{{\rm N} I}$ are 
the Newtonian location and velocity, respectively, 
$\boldsymbol{n}_I \equiv \boldsymbol{r}_{{\rm N} I} / r_{{\rm N} I}$, and 
$\boldsymbol{n}_{\perp I} \equiv \boldsymbol{v}_{{\rm N} I} / v_{{\rm N} I}$.
Note that $\boldsymbol{n}_I \perp \boldsymbol{n}_{\perp I}$ 
in the circular motion.
The PN term $\boldsymbol{\delta}_{{\rm EIH} 1}$ is defined by
\begin{align}
\boldsymbol{\delta}_{{\rm EIH} 1} &= 
\frac{1}{16}\frac{M}{\ell^2} \frac{1}{\sqrt{\nu_2^2 + \nu_2\nu_3 + \nu_3^2} }
\notag\\
&~~~
\times
\biggl\{\{16 (\nu_2^2 + \nu_2\nu_3 + \nu_3^2)
[3 - (\nu_1\nu_2 + \nu_2\nu_3 + \nu_3\nu_1)]
\notag\\
&~~~
+ 9\nu_2\nu_3[2(\nu_2 + \nu_3) + \nu_2^2 + 4\nu_2\nu_3 + \nu_3^2]\}
\boldsymbol{n}_{1}
\notag\\
&~~~
+ 3\sqrt{3}\nu_2\nu_3(\nu_2 - \nu_3)(5 - 3\nu_1)
\boldsymbol{n}_{\perp 1}
\biggr\} .
\label{C5eq:delta}
\end{align}
One can obtain the equations of motion for $m_2$ and $m_3$ 
by cyclic manipulations as $1 \to 2 \to 3 \to 1$

The PN triangular configuration becomes 
an equilibrium solution in the circular motion 
if and only if the following conditions (a) and (b) simultaneously hold:
(a) the term proportional to $\bs{n}_{\perp I}$ vanishes
and 
(b) all the angular velocities are the same.

As a result, 
we obtain the PN corrections that satisfy the above conditions as 
\cite{YA3}
\begin{align}
\rho_{12} &= \frac{1}{24}
[(\nu_2 - \nu_3)(5 - 3\nu_1) - (\nu_3 - \nu_1)(5 - 3\nu_2)] \varepsilon ,
\label{C5eq:d12}
\\
\rho_{23} &= \frac{1}{24}
[(\nu_3 - \nu_1)(5 - 3\nu_2) - (\nu_1 - \nu_2)(5 - 3\nu_3)] \varepsilon ,
\label{C5eq:d23}
\\
\rho_{31} &= \frac{1}{24}
[(\nu_1 - \nu_2)(5 - 3\nu_3) - (\nu_2 - \nu_3)(5 - 3\nu_1)] \varepsilon ,
\label{C5eq:d31}
\end{align}
which give a PN triangular equilibrium solution for general masses.
In this case, 
the angular velocities of the bodies are equal 
(i.e. $\omega_1 = \omega_2 = \omega_3 = \omega$).
These corrections for the restricted three-body problem as $\nu_3 \to 0$ 
reduce to previous results \cite{Krefetz, Maindl}.

Substituting Eq. \eqref{C5eq:d12} and Eq. \eqref{C5eq:d31} 
into Eq. \eqref{C5eq:M1-EOM-C}, 
we obtain the angular velocity of the three bodies
\begin{align}
\omega = \omega_{{\rm N}} ( 1 + \tilde{\omega}_{{\rm PN}} ),
\end{align}
where the Newtonian angular velocity $\omega_{{\rm N}} = \sqrt{M/\ell^3}$ and 
the dimensionless 1PN correction
\begin{align}
\tilde{\omega}_{{\rm PN}} = 
- \frac{1}{16} [29 - 14(\nu_1\nu_2 + \nu_2\nu_3 + \nu_3\nu_1)] \varepsilon .
\label{C5eq:omega-PN}
\end{align}
Using a relation $\nu_1 + \nu_2 + \nu_3 = 1$, 
we can show $\tilde{\omega}_{{\rm PN}} < 0$, that is, 
$\omega < \omega_{{\rm N}}$ for the fixed system parameters $\ell$ and $M$.
In other words, 
the PN triangular configuration is always smaller than the Newtonian one 
if the masses and angular velocity of the three bodies are fixed.

\section{The  Stability of the Post-Newtonian Triangular Solution}
\label{Stability}

Next, we study the stability of the PN triangular solution
by taking account of linear perturbations in the orbital plane.
It is convenient to use the corotating coordinates 
with the origin as the center of mass even after adding perturbations.
Therefore, 
the number of the degrees of freedom for the perturbations 
decreases from six to four.
One of them corresponds to 
a perturbation in the angular velocity, 
and the three other perturbations 
denote changes in the shape and size of the PN triangle.

We consider two perturbations 
corresponding to changes in the distances $r_{12}$ and $r_{31}$ as
\begin{align}
r_{12} = \ell (1 + \rho_{12} + \delta \chi_{12}) , \\
r_{31} = \ell (1 + \rho_{31} + \delta \chi_{31}) , 
\end{align}
where $\chi_{12}$ and $\chi_{31}$ are perturbations in the distances 
and $\delta$ is a {\it bookkeeping parameter} 
that denotes the smallness of the perturbations.
And, a perturbation $\psi_{23}$ denotes a change in 
the angle $\varphi_{23}$ between $r_{12}$ and $r_{31}$ as 
\begin{align}
\varphi_{23} = 
\frac{\pi}{3} + \sqrt{3} \rho_{23} + \delta \psi_{23} .
\end{align}
These three perturbations mean changes in
the shape and size of the PN triangle.
For the remaining one of the degrees of freedom 
corresponding to a change in the angular velocity, 
we denote 
\begin{align}
\theta_{12} = \Theta_{12} + \delta \sigma ,
\end{align}
where $\theta_{12}$ and $\sigma$ denote 
the direction of $\bs{r}_{12}$ to the reference frame and 
a perturbation in it, 
respectively.
$\Theta_{12}$ is the unperturbed direction, which 
satisfies the equation as
\begin{align}
\frac{d \Theta_{12}}{d t} = 
\omega_{\rm N} ( 1 + \tilde{\omega}_{\rm PN} ) .
\end{align}
Note that 
the 1PN corrections $\rho_{I J}$ satisfy Eq. \eqref{cc-1PNcorrection}, 
while the perturbations are arbitrary.
Figure \ref{C6fig:fig-triangle} shows 
a schematic figure for these four perturbations.
The above choice of the perturbations is convenient, 
because we can avoid directly using the PN center of mass 
\cite{MTW,LL,Will}.

Note that 
the perturbations $\chi_{12}$, $\chi_{31}$, $\psi_{23}$, and $\sigma$ have 
not only the Newtonian terms but also the 1PN ones. 
For instance, the perturbation $\sigma$ can be expanded as
\begin{align}
\sigma = \sigma_{\rm N} + \sigma_{\rm PN} , 
\end{align}
where $\sigma_{\rm N}$ and $\sigma_{\rm PN} (= \mathrm{O}(\varepsilon))$ are 
the Newtonian term and the 1PN one, respectively.
In the following, 
we neglect the terms of second (and higher) order in $\delta$.
Namely, we calculate 
to the terms of order $\varepsilon \times \delta$ 
(i.e. the linear perturbation at the 1PN order).

Defining a new variable as $X \equiv \chi_{31} - \chi_{12}$, 
we obtain the equation of motion for $\bs{r}_{12}$ 
(see Appendix \ref{app1} for a detailed derivation of the equations). 
Its radial part is
\begin{align}
&\left[
(D^2 - 3) \chi_{12} - 2 D \sigma 
- \frac94 \nu_3 X - \frac{3 \sqrt{3}}{4} \nu_3 \psi_{23} \right]
+ \varepsilon 
\Biggl[
- \frac{1}{32} \biggl\{
4 \sqrt{3} (\nu_1 - \nu_2) (7 - 9 \nu_3) \nu_3 D
\notag\\
&
+ (36 \nu_2^3 + 234 \nu_1 \nu_2^2 - 146 \nu_2^2 + 261 \nu_1^2 \nu_2 
- 488 \nu_1 \nu_2 + 155 \nu_2 + 63 \nu_1^3 - 155 \nu_1^2 + 137 \nu_1 
\notag\\
&
- 585) \biggr\} \chi_{12}
- \frac{1}{24} 
(27 \nu_2^3 + 135 \nu_1 \nu_2^2 - 21 \nu_2^2 + 135 \nu_1^2 \nu_2 
- 210 \nu_1 \nu_2 + 24 \nu_2 + 27 \nu_1^3 - 21 \nu_1^2 
\notag\\
&
+ 24 \nu_1- 155) D \sigma
- \frac{1}{32} \nu_3 \biggl\{
4 \sqrt{3} (9 \nu_1 \nu_2 + 10 \nu_2 + 9 \nu_1^2 - 6 \nu_1 - 4) D
- (216 \nu_2^2 + 288 \nu_1 \nu_2 
\notag\\
&
- 154 \nu_2 + 171 \nu_1^2 - 38 \nu_1 + 420)
\biggr\} X
+ \frac{1}{32} \nu_3 \biggl\{
4 (18 \nu_2^2 + 27 \nu_1 \nu_2 - 2 \nu_2 + 9 \nu_1^2 + 14 \nu_1 
\notag\\
&
- 12) D
+ \sqrt{3} 
(51 \nu_2^2 + 114 \nu_1 \nu_2 + 2 \nu_2 + 87 \nu_1^2 - 120 \nu_1 + 155)
\biggr\} \psi_{23}
\Biggr] = 0 ,
\label{C6eq:z12Re}
\end{align}
and the tangential part is 
\begin{align}
& \left[
2 D \chi_{12} + D^2 \sigma 
- \frac{3 \sqrt{3}}{4} \nu_3 X + \frac94 \nu_3 \psi_{23}
\right]
+ \varepsilon \Biggl[
- \frac{1}{32} \biggl\{
4 (9 \nu_2^3 + 45 \nu_1 \nu_2^2 + 9 \nu_2^2 + 45 \nu_1^2 \nu_2 
\notag\\
&
- 30 \nu_1 \nu_2 
- 18 \nu_2 + 9 \nu_1^3 + 9 \nu_1^2 - 18 \nu_1 + 61) D
+ 3 \sqrt{3} \nu_3 (12 \nu_2^2 - 6 \nu_1 \nu_2 + 14 \nu_2 - 15 \nu_1^2 
\notag\\
&
+ 4 \nu_1 - 5) \biggr\} \chi_{12}
- \frac{1}{24} \biggl\{
(3 \nu_2^2 + 12 \nu_1 \nu_2 - 18 \nu_2 + 3 \nu_1^2 - 18 \nu_1 + 10) D^2
- 3 \sqrt{3} (\nu_1 - \nu_2) 
\notag\\
&
\times \nu_3 (9 \nu_2 + 9 \nu_1 + 4) D
\biggr\} \sigma
+ \frac{1}{32} \nu_3 \biggl\{
4 (18 \nu_2^2 + 27 \nu_1 \nu_2 + 8 \nu_2 + 9 \nu_1^2 + 16 \nu_1 - 12) D
\notag\\
&
+ \sqrt{3} (36 \nu_2^2 + 72 \nu_1 \nu_2 - 54 \nu_2 + 81 \nu_1^2 - 90 \nu_1 + 160)
\biggr\} X
+ \frac{1}{32} \nu_3 \biggl\{
4 \sqrt{3} (9 \nu_1 \nu_2 + 8 \nu_2 
\notag\\
&
+ 9 \nu_1^2 - 4) D
- 9 (21 \nu_2^2 + 14 \nu_1 \nu_2 - 10 \nu_2 + 13 \nu_1^2 - 8 \nu_1 + 45)
\biggr\} \psi_{23}
\Biggr] = 0 ,
\label{C6eq:z12Im}
\end{align}
where $D$ denotes a differential operator with respect to
a normalized time $\tilde{t} \equiv \omega_{\rm N} t$.

In the same way, 
we obtain the equation of motion for $\bs{r}_{31}$ and its radial 
and tangential parts are
\begin{align}
& \left[
( D^2 - 3 ) \chi_{12} - 2 D \sigma + \left( D^2 - 3 + \frac94 \nu_2 \right) X 
- \left(2 D + \frac{3 \sqrt{3}}{4} \nu_2 \right) \psi_{23} \right]
+ \varepsilon \Biggl[
- \frac{1}{32} \biggl\{ 4 \sqrt{3} 
\notag\\
&
\times (\nu_3 - \nu_1) (7 - 9 \nu_2) \nu_2 D
+ (36 \nu_3^3 + 234 \nu_1 \nu_3^2 - 146 \nu_3^2 + 261 \nu_1^2 \nu_3 
- 488 \nu_1 \nu_3 + 155 \nu_3 
\notag\\
&
+ 63 \nu_1^3 - 155 \nu_1^2 + 137 \nu_1 - 585) 
\biggr\} \chi_{12}
- \frac{1}{24} 
(27 \nu_3^3 + 135 \nu_1 \nu_3^2 - 21 \nu_3^2 + 135 \nu_1^2 \nu_3 
- 210 \nu_1 \nu_3 
\notag\\
&
+ 24 \nu_3 + 27 \nu_1^3 - 21 \nu_1^2 + 24 \nu_1 - 155) D \sigma
- \frac{1}{32} \biggl\{ 
4 \sqrt{3} \nu_2 (9 \nu_3^2 + 9 \nu_1 \nu_3 + 8 \nu_3 - 4 \nu_1 - 4) D
\notag\\
&
- (180 \nu_3^3 + 270 \nu_1 \nu_3^2 - 224 \nu_3^2 + 198 \nu_1^2 \nu_3 
+ 8 \nu_1 \nu_3 + 419 \nu_3 + 108 \nu_1^3 - 54 \nu_1^2 + 321 \nu_1 
\notag\\
&
+ 165)
\biggr\} X
+ \frac{1}{96} \biggl\{ 
4 (27 \nu_3^3 - 39 \nu_3^2 - 27 \nu_1^2 \nu_3 + 165 \nu_1 \nu_3 - 54 \nu_3 
+ 36 \nu_1^2 - 102 \nu_1 + 191) D
\notag\\
&
+ 3 \sqrt{3} \nu_2 
(51 \nu_3^2 + 114 \nu_1 \nu_3 + 2 \nu_3 + 87 \nu_1^2 - 120 \nu_1 + 155)
\biggr\} \psi_{23}
\Biggr] = 0 ,
\label{C6eq:z31Re}
\end{align}

\begin{align}
& \left[
2 D \chi_{12} + D^2 \sigma + \left(2 D - \frac{3 \sqrt{3}}{4} \nu_2 \right) X 
+ \left(D^2 - \frac94 \nu_2 \right) \psi_{23}
\right]
+ \varepsilon \Biggl[
- \frac{1}{32} \biggl\{
4 (9 \nu_3^3 + 45 \nu_1 \nu_3^2 
\notag\\
&
+ 9 \nu_3^2 + 45 \nu_1^2 \nu_3 - 30 \nu_1 \nu_3 
- 18 \nu_3 + 9 \nu_1^3 + 9 \nu_1^2 - 18 \nu_1 + 61) D
- 3 \sqrt{3} \nu_2 (12 \nu_3^2 - 6 \nu_1 \nu_3 
\notag\\
&
+ 14 \nu_3 - 15 \nu_1^2 
+ 4 \nu_1 - 5)
\biggr\} \chi_{12} 
- \frac{1}{24} \biggl\{
(3 \nu_3^2 + 12 \nu_1 \nu_3 - 18 \nu_3 + 3 \nu_1^2 - 18 \nu_1 + 10) D^2
\notag\\
&
- 3 \sqrt{3} (\nu_3 - \nu_1) (13 - 9 \nu_2) \nu_2 D
\biggr\} \sigma 
+ \frac{1}{32} \biggl\{
4 (9 \nu_3^3 - 19 \nu_3^2 - 9 \nu_1^2 \nu_3 + 27 \nu_1 \nu_3 - 2 \nu_3 
- 2 \nu_1^2 
\notag\\
&
- 10 \nu_1 - 49) D
+ \sqrt{3} 
(72 \nu_3^2 + 54 \nu_1 \nu_3 - 12 \nu_3 + 36 \nu_1^2 - 78 \nu_1 + 145) \nu_2
\biggr\} X 
- \frac{1}{96} \biggl\{
4 (3 \nu_3^2 
\notag\\
&
+ 12 \nu_1 \nu_3 - 18 \nu_3 + 3 \nu_1^2 - 18 \nu_1 + 10) D^2
- 12 \sqrt{3} (9 \nu_3^2 + 9 \nu_1 \nu_3 + 12 \nu_3 - 4 \nu_1 - 4) \nu_2 D
\notag\\
&
- 27 (21 \nu_3^2 + 14 \nu_1 \nu_3 - 10 \nu_3 + 13 \nu_1^2 - 8 \nu_1 + 45) \nu_2
\biggr\} \psi_{23}
\Biggr] = 0 .
\label{C6eq:z31Im}
\end{align}

First, we study the condition for stability in Newtonian gravity.
In the Newtonian limit $\varepsilon \to 0$, 
the equations of motion \eqref{C6eq:z12Re} - \eqref{C6eq:z31Im} 
for the perturbations are rearranged as 
\begin{align}
\begin{pmatrix}
(D^2 - 3) & - 2 D & \displaystyle{- \frac94 \nu_3} & 
\displaystyle{- \frac{3 \sqrt{3}}{4} \nu_3} \\
2 D & D^2 & \displaystyle{- \frac{3 \sqrt{3}}{4} \nu_3} & 
\displaystyle{\frac94 \nu_3} \\
( D^2 - 3 ) & - 2 D & \displaystyle{D^2 - 3 + \frac94 \nu_2} & 
\displaystyle{- \left(2 D + \frac{3 \sqrt{3}}{4} \nu_2 \right)} \\
2 D & D^2 & \displaystyle{2 D - \frac{3 \sqrt{3}}{4} \nu_2} & 
\displaystyle{D^2 - \frac94 \nu_2} 
\end{pmatrix}
\begin{pmatrix}
\chi_{12} \\[10pt] \sigma \\[10pt] X \\[10pt] \psi_{23} 
\end{pmatrix}
= 0 .
\label{C6eq:EoM-perturbation-Newton}
\end{align}
Carrying out a classical stability analysis (e.g. \cite{Danby}), 
we obtain the eigenvalue equation, so called the secular equation as
\begin{align}
\lambda^2 ( \lambda^2 + 1 ) 
\left( \lambda^2 + \frac{1 + \sqrt{1 - 27 V}}{2} \right)
\left( \lambda^2 + \frac{1 - \sqrt{1 - 27 V}}{2} \right)
= 0 ,
\label{C6eq:Secular-eq-Newton}
\end{align}
where $\lambda$ is the eigenvalue and 
$V \equiv \nu_1 \nu_2 + \nu_2 \nu_3 + \nu_3 \nu_1$.
Since Eq. \eqref{C6eq:Secular-eq-Newton} has only 
the terms of the even-order of $\lambda$, 
if $\operatorname{Re} (\lambda) \neq 0$, 
both the positive and negative real parts are allowed 
and then the equilibrium configuration is unstable.
In fact, 
the roots of Eq. \eqref{C6eq:Secular-eq-Newton} are 
\begin{align}
\lambda_0 = 0 , \,\,
\lambda_{1 \pm} =  \pm i , \,\,  
\lambda_{2 \pm} = \pm \sqrt{- \frac{1 + \sqrt{1 - 27 V}}{2}} , \,\, 
\lambda_{3 \pm} = \pm \sqrt{- \frac{1 - \sqrt{1 - 27 V}}{2}} .
\label{C6eq:roots-Newton}
\end{align}
Therefore, 
Lagrange's equilateral triangular solution is stable 
if and only if 
$\lambda_{2 \pm}$ and $\lambda_{3 \pm}$ are purely imaginary.
Hence,
it is necessary and sufficient that 
\begin{align}
1 - 27 V > 0 .
\end{align}
This is nothing but the Newtonian condition Eq. \eqref{CoSN} for stability 
of Lagrange's solution.

Next, let us consider at the 1PN order.
The EIH equations of motion 
\eqref{C6eq:z12Re} - \eqref{C6eq:z31Im} for the perturbations are
\begin{align}
\begin{pmatrix}
M_{11} & M_{12} & M_{13} & M_{14} \\
M_{21} & M_{22} & M_{23} & M_{24} \\
M_{31} & M_{32} & M_{33} & M_{34} \\
M_{41} & M_{42} & M_{43} & M_{44} 
\end{pmatrix}
\begin{pmatrix}
\chi_{12} \\ \sigma \\ X \\ \psi_{23} 
\end{pmatrix}
= 0 ,
\label{C6eq:EoM-perturbation-Matrix}
\end{align}
where 
\begin{align}
M_{11} &= (D^2 - 3) 
- \frac{1}{32} \varepsilon \Bigl\{
4 \sqrt{3} (\nu_1 - \nu_2) (7 - 9 \nu_3) \nu_3 D
+ (36 \nu_2^3 + 234 \nu_1 \nu_2^2 - 146 \nu_2^2 
\notag\\
&~~~
+ 261 \nu_1^2 \nu_2 - 488 \nu_1 \nu_2 + 155 \nu_2 + 63 \nu_1^3 - 155 \nu_1^2 
+ 137 \nu_1 - 585) \Bigr\} , \\
M_{12} &= - 2 D
- \frac{1}{24} \varepsilon
(27 \nu_2^3 + 135 \nu_1 \nu_2^2 - 21 \nu_2^2 + 135 \nu_1^2 \nu_2 
- 210 \nu_1 \nu_2 + 24 \nu_2 + 27 \nu_1^3 
\notag\\
&~~~
- 21 \nu_1^2 + 24 \nu_1- 155) D , \\
M_{13} &= - \frac94 \nu_3
- \frac{1}{32} \varepsilon \nu_3 \Bigl\{
4 \sqrt{3} (9 \nu_1 \nu_2 + 10 \nu_2 + 9 \nu_1^2 - 6 \nu_1 - 4) D
- (216 \nu_2^2 + 288 \nu_1 \nu_2 
\notag\\
&~~~
- 154 \nu_2 + 171 \nu_1^2 - 38 \nu_1 + 420) \Bigr\} , \\
M_{14} &= - \frac{3 \sqrt{3}}{4} \nu_3
+ \frac{1}{32} \varepsilon \nu_3 \Bigl\{
4 (18 \nu_2^2 + 27 \nu_1 \nu_2 - 2 \nu_2 + 9 \nu_1^2 + 14 \nu_1 - 12) D
+ \sqrt{3} (51 \nu_2^2 
\notag\\
&~~~
+ 114 \nu_1 \nu_2 + 2 \nu_2 + 87 \nu_1^2 - 120 \nu_1 + 155) \Bigr\} , 
\\
M_{21} &= 2 D
- \frac{1}{32} \varepsilon \Bigl\{
4 (9 \nu_2^3 + 45 \nu_1 \nu_2^2 + 9 \nu_2^2 + 45 \nu_1^2 \nu_2 
- 30 \nu_1 \nu_2 - 18 \nu_2 + 9 \nu_1^3 + 9 \nu_1^2 
\notag\\
&~~~
- 18 \nu_1 + 61) D
+ 3 \sqrt{3} \nu_3 (12 \nu_2^2 - 6 \nu_1 \nu_2 + 14 \nu_2 - 15 \nu_1^2 
+ 4 \nu_1 - 5) \Bigr\} , \\
M_{22} &= D^2 
- \frac{1}{24} \varepsilon \Bigl\{
(3 \nu_2^2 + 12 \nu_1 \nu_2 - 18 \nu_2 + 3 \nu_1^2 - 18 \nu_1 + 10) D^2
- 3 \sqrt{3} (\nu_1 - \nu_2) \nu_3 
\notag\\
&~~~
\times (9 \nu_2 + 9 \nu_1 + 4) D \Bigr\} , \\
M_{23} &= - \frac{3 \sqrt{3}}{4} \nu_3
+ \frac{1}{32} \varepsilon \nu_3 \Bigl\{
4 (18 \nu_2^2 + 27 \nu_1 \nu_2 + 8 \nu_2 + 9 \nu_1^2 + 16 \nu_1 - 12) D
+ \sqrt{3} (36 \nu_2^2 
\notag\\
&~~~
+ 72 \nu_1 \nu_2 - 54 \nu_2 + 81 \nu_1^2 - 90 \nu_1 + 160) \Bigr\} , 
\end{align}

\begin{align}
M_{24} &= \frac94 \nu_3
+ \frac{1}{32} \varepsilon \nu_3 \Bigl\{
4 \sqrt{3} (9 \nu_1 \nu_2 + 8 \nu_2 
+ 9 \nu_1^2 - 4) D
- 9 (21 \nu_2^2 + 14 \nu_1 \nu_2 - 10 \nu_2 
\notag\\
&~~~
+ 13 \nu_1^2 - 8 \nu_1 + 45) \Bigr\} ,
\\
M_{31} &= ( D^2 - 3 )
- \frac{1}{32} \varepsilon \Bigl\{ 
4 \sqrt{3} (\nu_3 - \nu_1) (7 - 9 \nu_2) \nu_2 D
+ (36 \nu_3^3 + 234 \nu_1 \nu_3^2 - 146 \nu_3^2 
\notag\\
&~~~
+ 261 \nu_1^2 \nu_3 
- 488 \nu_1 \nu_3 + 155 \nu_3 + 63 \nu_1^3 - 155 \nu_1^2 + 137 \nu_1 - 585) 
\Bigr\} , \\
M_{32} &= - 2 D
- \frac{1}{24} \varepsilon 
(27 \nu_3^3 + 135 \nu_1 \nu_3^2 - 21 \nu_3^2 + 135 \nu_1^2 \nu_3 
- 210 \nu_1 \nu_3 + 24 \nu_3 + 27 \nu_1^3 
\notag\\
&~~~
- 21 \nu_1^2 + 24 \nu_1 - 155) D , \\
M_{33} &= D^2 - 3 + \frac94 \nu_2
- \frac{1}{32} \varepsilon \Bigl\{ 
4 \sqrt{3} \nu_2 (9 \nu_3^2 + 9 \nu_1 \nu_3 + 8 \nu_3 - 4 \nu_1 - 4) D
- (180 \nu_3^3 + 270 \nu_1 \nu_3^2 
\notag\\
&~~~
- 224 \nu_3^2 + 198 \nu_1^2 \nu_3 
+ 8 \nu_1 \nu_3 + 419 \nu_3 + 108 \nu_1^3 - 54 \nu_1^2 + 321 \nu_1 + 165)
\Bigr\} , 
\\
M_{34} &= - \left(2 D + \frac{3 \sqrt{3}}{4} \nu_2 \right)
+ \frac{1}{96} \varepsilon \Bigl\{ 
4 (27 \nu_3^3 - 39 \nu_3^2 - 27 \nu_1^2 \nu_3 + 165 \nu_1 \nu_3 - 54 \nu_3 
+ 36 \nu_1^2 
\notag\\
&~~~
- 102 \nu_1 + 191) D
+ 3 \sqrt{3} \nu_2 
(51 \nu_3^2 + 114 \nu_1 \nu_3 + 2 \nu_3 + 87 \nu_1^2 - 120 \nu_1 + 155) \Bigr\} ,
\\
M_{41} &= 2 D
- \frac{1}{32} \varepsilon \Bigl\{
4 (9 \nu_3^3 + 45 \nu_1 \nu_3^2 + 9 \nu_3^2 + 45 \nu_1^2 \nu_3 - 30 \nu_1 \nu_3 
- 18 \nu_3 + 9 \nu_1^3 + 9 \nu_1^2 
\notag\\
&~~~
- 18 \nu_1 + 61) D
- 3 \sqrt{3} \nu_2 (12 \nu_3^2 - 6 \nu_1 \nu_3 + 14 \nu_3 - 15 \nu_1^2 
+ 4 \nu_1 - 5) \biggr\} , \\
M_{42} &= D^2
- \frac{1}{24} \varepsilon \Bigl\{
(3 \nu_3^2 + 12 \nu_1 \nu_3 - 18 \nu_3 + 3 \nu_1^2 - 18 \nu_1 + 10) D^2
- 3 \sqrt{3} (\nu_3 - \nu_1) 
\notag\\
&~~~
\times (13 - 9 \nu_2) \nu_2 D \Bigr\} , \\
M_{43} &= 2 D - \frac{3 \sqrt{3}}{4} \nu_2
+ \frac{1}{32} \varepsilon \Bigl\{
4 (9 \nu_3^3 - 19 \nu_3^2 - 9 \nu_1^2 \nu_3 + 27 \nu_1 \nu_3 - 2 \nu_3 
- 2 \nu_1^2 - 10 \nu_1 - 49) D
\notag\\
&~~~
+ \sqrt{3} 
(72 \nu_3^2 + 54 \nu_1 \nu_3 - 12 \nu_3 + 36 \nu_1^2 - 78 \nu_1 + 145) \nu_2
\Bigr\} , 
\\
M_{44} &= D^2 - \frac94 \nu_2
- \frac{1}{96} \varepsilon \Bigl\{
4 (3 \nu_3^2 + 12 \nu_1 \nu_3 - 18 \nu_3 + 3 \nu_1^2 - 18 \nu_1 + 10) D^2
- 12 \sqrt{3} (9 \nu_3^2 + 9 \nu_1 \nu_3 
\notag\\
&~~~
+ 12 \nu_3 - 4 \nu_1 - 4) \nu_2 D
- 27 (21 \nu_3^2 + 14 \nu_1 \nu_3 - 10 \nu_3 + 13 \nu_1^2 - 8 \nu_1 + 45) \nu_2
\Bigr\} .
\end{align}

In a similar manner to the Newtonian case, 
we obtain the secular equation at the 1PN order as
\begin{align}
\lambda^2 \Biggl[
\lambda^6 + 2 \left\{ 
1 - \frac18 \varepsilon ( 77 - 10 V ) \right\} \lambda^4
+ \left\{ 1 + \frac{27}{4} V 
- \frac{1}{16} \varepsilon ( 308 + 1265 V + 162 W - 378 V^2 ) \right\} \lambda^2
\notag\\
+ \frac{27}{4} \left\{ V - \frac{1}{24} \varepsilon 
( 521 V - 72 W - 126 V^2 ) \right\} \Biggr] = 0 ,
\label{C6eq:Secular-original}
\end{align}
where $W \equiv \nu_1 \nu_2 \nu_3$.

Neglecting the trivial root $\lambda = 0$, 
we obtain a cubic equation of $\tau \equiv \lambda^2$ as
\begin{align}
\tau^3 + \alpha \tau^2 + \beta \tau + \gamma = 0 ,
\label{C6eq:Secular-eq}
\end{align}
where
\begin{align}
\alpha &\equiv 
2 \left\{ 1 - \frac18 \varepsilon \left( 77 - 10 V \right) \right\} , \\
\beta &\equiv 
1 + \frac{27}{4} V 
- \frac{1}{16} \varepsilon \left( 308 + 1265 V + 162 W - 378 V^2 \right) , \\
\gamma &\equiv 
\frac{27}{4} 
\left\{ V - \frac{1}{24} \varepsilon ( 521 V - 72 W - 126 V^2 ) \right\} .
\end{align}
In a similar manner to
the Newtonian case, 
the PN triangular solution is stable 
if and only if 
all the roots $\tau$ of Eq. \eqref{C6eq:Secular-eq} are negative real, 
so that all $\lambda = \pm \sqrt{\tau}$ have no real part.

\begin{figure}[ht]
\begin{center}
  \includegraphics[width=9cm]{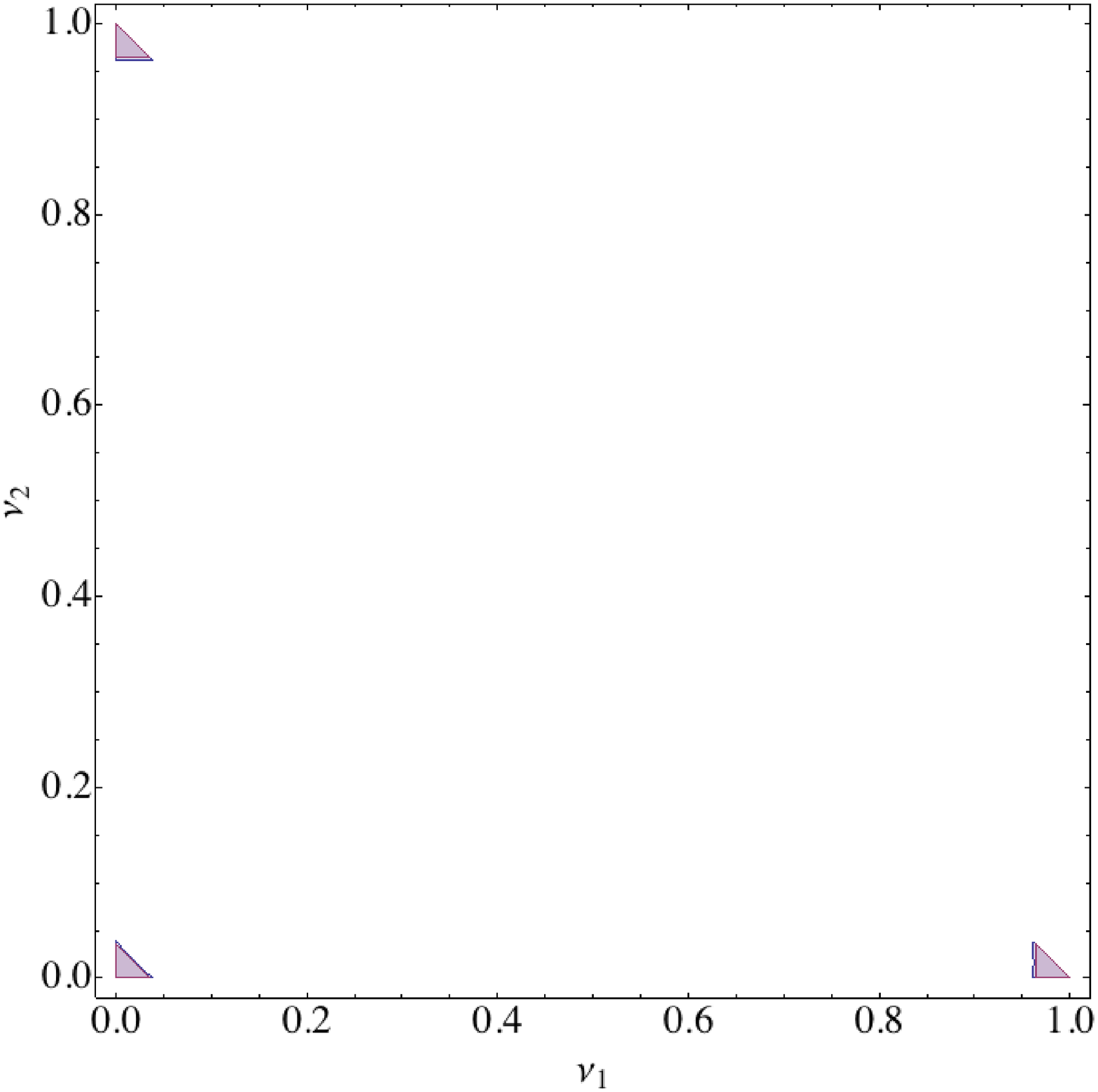}
  \includegraphics[width=9cm]{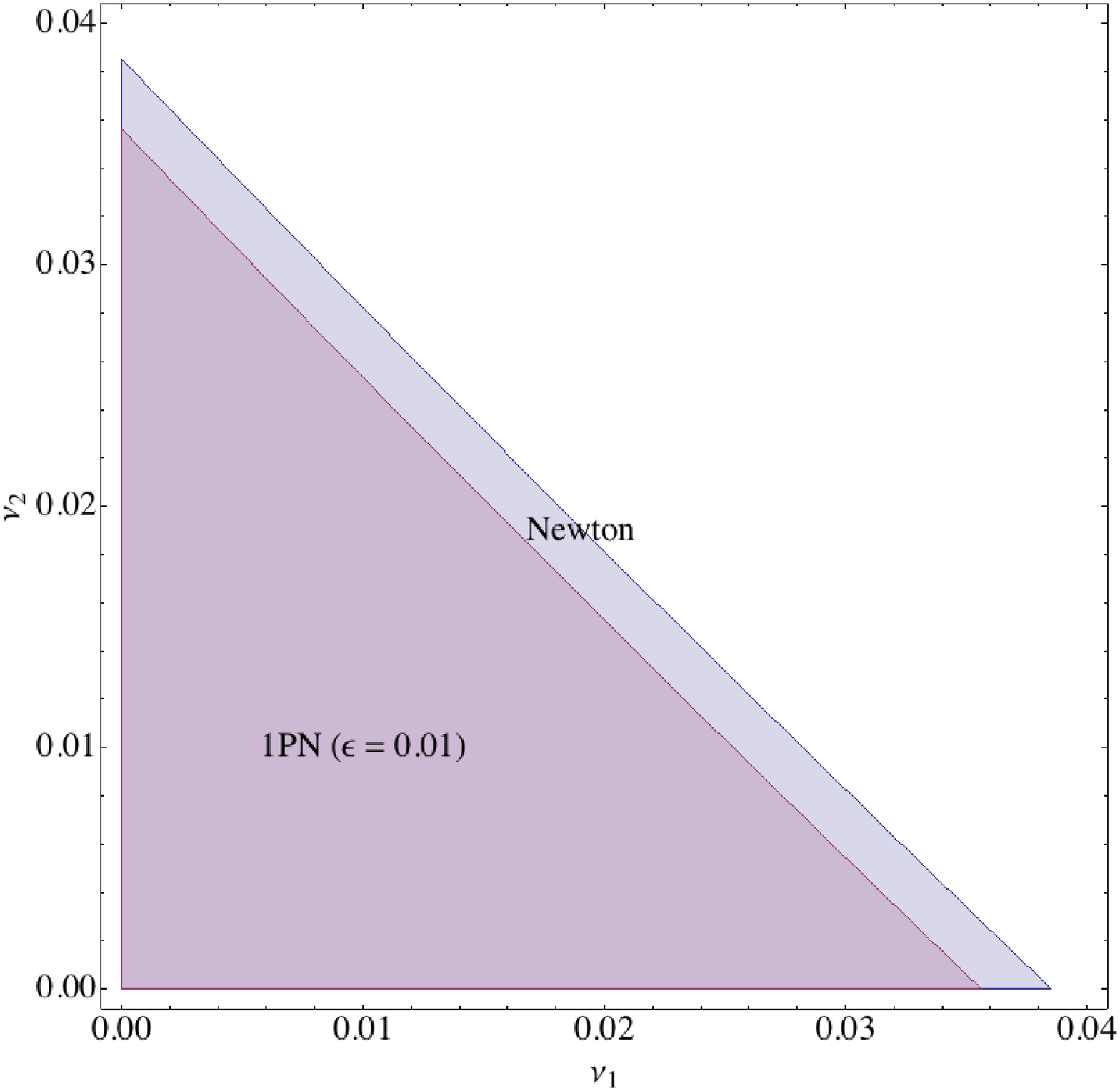}
\caption{
Mass ratios for a stable configuration that satisfies
the Newtonian condition Eq. \eqref{CoSN} and 
the PN one Eq. \eqref{C6eq:discriminant-2-1PN} 
when $\varepsilon = 0.01$ 
(i.e. the order of magnitude of the 1PN effects is 0.01), for instance.
For values of the mass ratios within the colored areas, 
the triangular configuration for three finite masses is stable.
Top: All the stability regions.
Bottom: The regions around small $\nu_1$ and $\nu_2$, 
where the third mass is dominant.
The stability regions at the 1PN order still exist, 
though they are more narrow than the Newtonian case.
}
\label{C6fig:region-1}
\end{center}
\end{figure}

In the 1PN approximation, 
the PN corrections to the Newtonian roots must be small.
Thus, we can factor Eq. \eqref{C6eq:Secular-eq} as
\begin{align}
(\tau + 1 - a \varepsilon )(\tau^2 + b \tau + c ) = 0 , 
\label{C6eq:Secular-factored}
\end{align}
where $a$, $b$, and $c$ are constants and the 2PN terms are neglected.
From Eqs. \eqref{C6eq:Secular-eq} and \eqref{C6eq:Secular-factored}, 
one can obtain
\begin{align}
a &= \frac{1}{8 V} ( 77 V - 14 V^2 - 36 W) , \\
b &= 1 - \frac{1}{8 V} ( 77 V - 6 V^2 + 36 W ) \varepsilon , 
\label{coef-b} \\
c &= \frac{27}{4} V - \frac{1}{16} ( 1305 V - 378 V^2 + 162 W) \varepsilon .
\label{coef-c}
\end{align}
Since $\varepsilon \ll 1$, 
we have $- 1 + a \varepsilon < 0$, $b > 0$, and $c>0$.
The roots of Eq. \eqref{C6eq:Secular-factored} are expressed as
\begin{align}
\tau_1 = - 1 + a \varepsilon, \,\,
\tau_{\pm} = \frac{- b \pm \sqrt{b^2 - 4 c}}{2} .
\label{C5eq:root-QE}
\end{align}
Hence, 
Eq. \eqref{C6eq:Secular-factored} has three negative real roots 
if and only if 
all the roots \eqref{C5eq:root-QE} are negative real.
Namely,
it is necessary and sufficient for the stability that 
\begin{align}
b^2 - 4 c > 0 .
\label{CoS1PN-o}
\end{align}
For the critical value $b^2 - 4 c = 0$, 
$V$ in the PN terms of Eq. \eqref{CoS1PN-o} 
with Eqs. \eqref{coef-b} and \eqref{coef-c}
can be replaced by the Newtonian critical value as
$V = 1/27$, 
because 1PN corrections to it make 2PN (or higher-order) contributions 
and they can be neglected.
Therefore, 
the condition for stability of the PN triangular solution becomes
\begin{align}
1 - \frac{391}{54} \varepsilon 
- 27 \left( V + \frac{15}{2} W \varepsilon \right) > 0 .
\end{align}
This is explicitly rewritten as
\begin{align}
\frac{m_1 m_2 + m_2 m_3 + m_3 m_1}{(m_1 + m_2 + m_3)^2} 
+ \frac{15}{2} \frac{m_1 m_2 m_3}{(m_1 + m_2 + m_3)^3}  \varepsilon < 
\frac{1}{27} \left( 1  - \frac{391}{54} \varepsilon \right) .
\label{C6eq:discriminant-2-1PN}
\end{align}
Equation \eqref{C6eq:discriminant-2-1PN} recovers 
the Newtonian condition \eqref{CoSN} in the limit $\varepsilon \to 0$.
The PN correction in the right-hand side of 
Eq. \eqref{C6eq:discriminant-2-1PN} is negative 
and 
the PN term of the triple product of masses 
in the left-hand side of Eq. \eqref{C6eq:discriminant-2-1PN} is positive.
Hence, the PN condition for stability is 
tighter than the Newtonian one 
for any positive small value of the parameter $\varepsilon$.
Figure \ref{C6fig:region-1} shows 
the Newtonian stability regions Eq. \eqref{CoSN} 
and the 1PN ones Eq. \eqref{C6eq:discriminant-2-1PN} 
when $\varepsilon = 0.01$ 
(i.e. the order of magnitude of the PN effects is $0.01$), for instance.

\begin{figure}[ht]
\begin{center}
  \includegraphics[width=10cm]{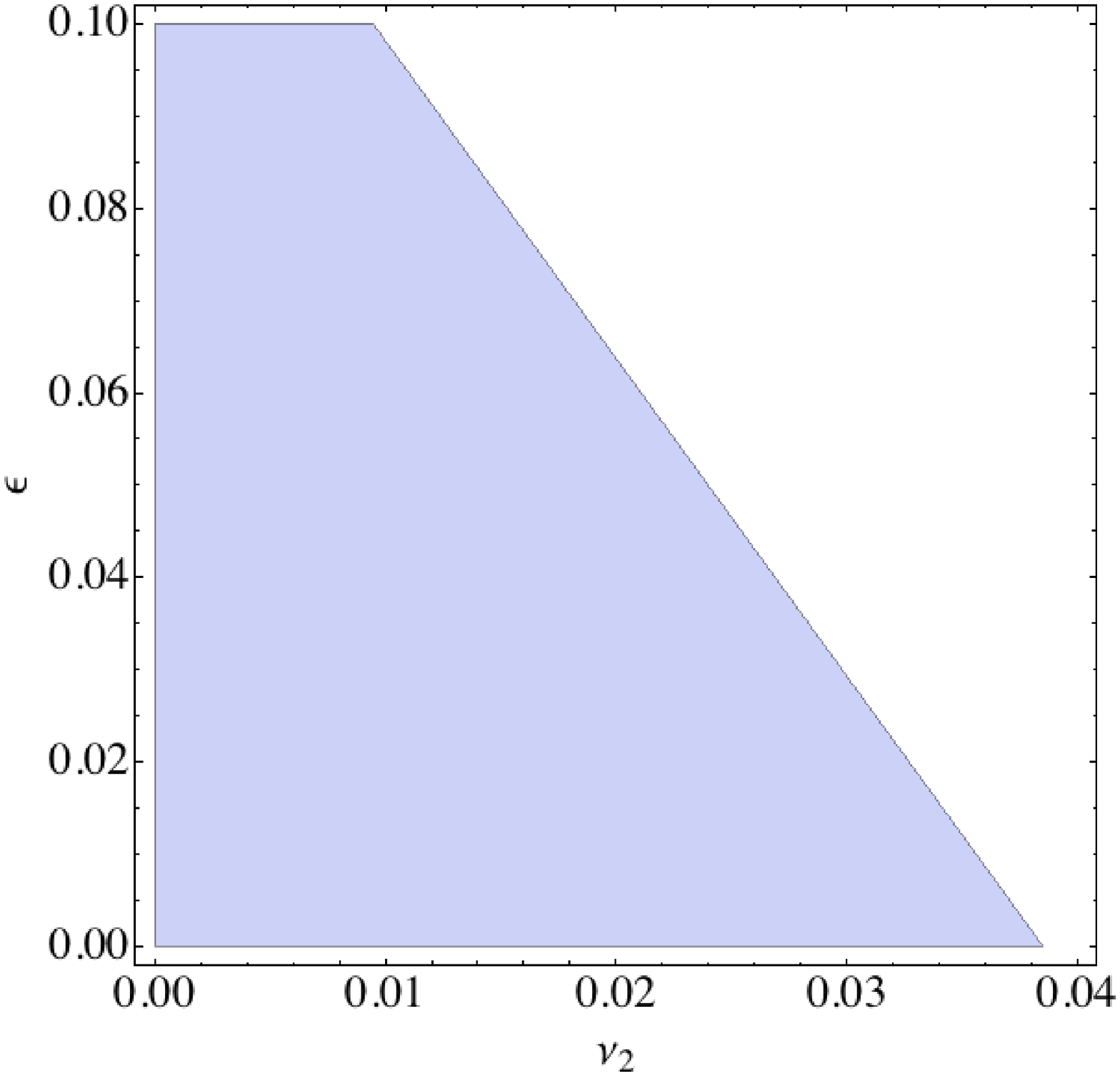}
\caption{
The stability region Eq. \eqref{C6eq:condition-R3BP-1PN} 
for the PN restricted three-body systems 
in the $(\nu_2, \varepsilon)$ plane.
For values of $\nu_2$ and $\varepsilon$ within the colored area, 
the PN triangular configuration in the restricted three-body case is stable.
This figure corresponds to Fig. 1 in Ref. \cite{DP}.
}
\label{C6fig:region-2}
\end{center}
\end{figure}

Finally, 
we focus on the restricted three-body limit as $\nu_3 \to 0$ (i.e. $W \to 0$).
In this case,
the stability condition Eq. \eqref{CoS1PN-o} becomes
\begin{align}
1 - 27 V - \frac{77 - 1311 V + 378 V^2}{4} \varepsilon > 0 .
\label{ineq-V}
\end{align}
This is a quadratic inequality of $V$. 
Solving Eq. \eqref{ineq-V} for $V$,
we obtain
\begin{align}
\frac{m_1 m_2}{(m_1 + m_2)^2} < 
\frac{1}{27} \left( 1 - \frac{391}{54} \varepsilon \right) ,
\label{ineq-m1m2}
\end{align}
where we used a relation 
$0 \leq V \leq 1/4$ in the restricted three-body problem.
Using a relation $\nu_1 + \nu_2 = 1$ and 
assuming $\nu_1 > \nu_2$ without loss of generality, 
we can rewrite Eq. \eqref{ineq-m1m2} as
\begin{align}
\frac{m_2}{m_1 + m_2} &< \mu_0 - \frac{17 \sqrt{69}}{486} \varepsilon , 
\label{C6eq:condition-R3BP-1PN}
\end{align}
where the Newtonian value $\mu_0 = (9 - \sqrt{69})/18$.
This condition is in agreement with previous results \cite{DP,SB1,SB2} 
(see Eq. \eqref{CoS-R3BP} in the present paper).
Figure \ref{C6fig:region-2} shows 
a region of stability in the PN restricted three-body problem.

\section{Conclusion}
\label{Conclusion}

We examined the PN effects on the stability of 
the triangular solution 
in the relativistic three-body problem for general masses.
The stability regions still exist even at the 1PN order.
The PN stability condition Eq. \eqref{C6eq:discriminant-2-1PN} 
is consistent with the Newtonian one Eq. \eqref{CoSN} 
in the limit as $\varepsilon \to 0$.
The PN correction in the right-hand side of 
Eq. \eqref{C6eq:discriminant-2-1PN}, 
which is in agreement with that in the PN restricted three-body problem 
\cite{DP,SB1,SB2}, 
makes the condition more strict than the Newtonian case 
for any small positive value of the parameter $\varepsilon$.

The PN term of the triple product of masses 
in the left-hand side of Eq. \eqref{C6eq:discriminant-2-1PN} 
appears not in the restricted case but in the general one.
The instability is also enhanced by this term, 
while this effect is smaller than the other PN one 
in the case of mass ratios for stable configurations.
If a system is mildly relativistic as $\varepsilon = 0.01$, 
for instance, 
the maximum value of $W$ is $\mathrm{O} (10^{-4})$ 
when $\nu_2 = \nu_3 \approx 0.019$ in a stability region.
Namely, 
the contribution from $W$ is comparable to the 2PN (or more higher) order.
This implies that 
triple systems with the PN triangular configuration for three finite masses 
are possible as well as restricted three-body systems.

The PN triangular configuration ought to emit gravitational waves 
\cite{Asada, THA}.
Such a system will shrink by gravitational radiation reaction 
if its configuration is initially stable, 
and the PN effects on the long-term stability should be incorporated.
In addition, we concentrate on the circular orbit in this paper.
Gravitational radiation is known to decrease eccentricity in binary orbits. 
It is left as a future work 
to study gravitational radiation reaction to the PN triangular configuration.

\section*{Acknowledgments}

We would like to thank Taihei Yano and Haruo Yoshida 
for providing with useful information on the literature.
We are grateful to Yuuiti Sendouda for useful comments.
This work was supported in part by JSPS Grant-in-Aid for JSPS Fellows, 
No. 24108 (K.Y.), 
and JSPS Grant-in-Aid for Scientific Research (Kiban C), No. 26400262 (H.A.).

\appendix
\section{A Derivation of the Equations of Motion for Perturbations}
\label{app1}

\begin{figure}[ht]
\begin{center}
  \includegraphics[width=10cm]{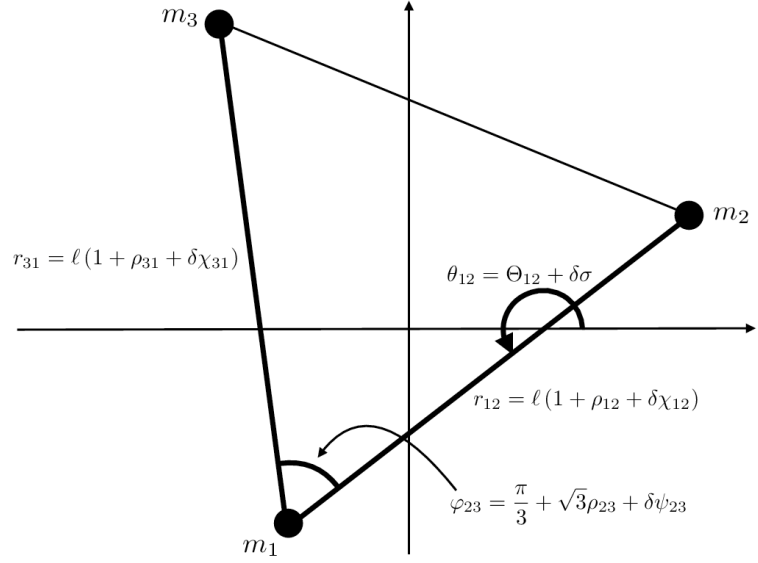}
\caption{Four perturbations in the PN triangular configuration. 
The perturbations $\chi_{12}$ and $\chi_{31}$ denote 
changes in $r_{12}$ and $r_{31}$, respectively, 
$\psi_{23}$ is a perturbation in the interior angle $\varphi_{23}$, 
and $\sigma$ corresponds to a change in the angular velocity.
}
\label{C6fig:fig-triangle}
\end{center}
\end{figure}

We consider four perturbations in the orbital plane
(see Fig. \ref{C6fig:fig-triangle}).
First, we put the distances between the bodies as
\begin{align}
r_{I J} = \ell (1 + \rho_{I J} + \delta \chi_{I J}) ,
\label{C6eq:separation}
\end{align}
where 
$\chi_{I J} ( = \chi_{J I} )$ is a perturbation in the distance $r_{I J}$
and $\delta$ is a bookkeeping parameter
that denotes the smallness of the perturbations.
By these perturbations, 
each angle $\varphi_{I J}$ between $\bs{r}_{K I}$ and $\bs{r}_{J K}$ 
($I \neq J \neq K$) of the PN triangle is changed as
\begin{align}
\varphi_{I J} = 
\frac{\pi}{3} + \sqrt{3} \rho_{I J} + \delta \psi_{I J} .
\label{C6eq:interior-angle}
\end{align}
The perturbations $\chi_{I J}$ and $\psi_{I J}$ relate to each other 
through the cosine formula, 
and the number of independent perturbations is three.

The remaining one of the degrees of freedom corresponds to 
a change in the angular velocity of the bodies:
\begin{align}
\theta_{I J} = \Theta_{I J} + \delta \sigma_{I J} ,
\end{align}
where $\theta_{I J}$ and $\sigma_{I J}$ denote 
the direction of $\bs{r}_{I J}$ to the reference frame and 
a perturbation in it, respectively.
$\Theta_{I J}$ is the unperturbed direction which satisfies the equation as
\begin{align}
\frac{d \Theta_{I J}}{d t} = 
\omega_{\rm N} ( 1 + \tilde{\omega}_{\rm PN} ) .
\end{align}
Differentiating relations as 
\begin{align}
\theta_{23} = \theta_{12} - \pi - \varphi_{31} , \\
\theta_{31} = \theta_{12} + \pi + \varphi_{23} , 
\end{align}
we obtain 
\begin{align}
D \sigma_{23} &= D (\sigma - \psi_{31}) , 
\label{ap:s23} \\
D \sigma_{31} &= D (\sigma + \psi_{23}) , 
\label{ap:s31} 
\end{align}
where 
$D$ denotes a differential operator with respect to
a normalized time $\tilde{t} \equiv \omega_{\rm N} t$ and
we denote $\sigma_{12}$ simply as $\sigma$.
Thus, the number of degrees of freedom for 
($\sigma_{12}, \sigma_{23}, \sigma_{31}$) is one 
and it corresponds to a change in the angular velocity.

Note that 
the perturbations have not only the Newtonian terms but also the 1PN ones.
For instance, the perturbation $\sigma$ can be expanded as
\begin{align}
\sigma = \sigma_{\rm N} + \sigma_{\rm PN} , 
\end{align}
where $\sigma_{\rm N}$ and $\sigma_{\rm PN} (= \mathrm{O}(\varepsilon))$ are 
the Newtonian term and the 1PN one, respectively.
In the following, 
we neglect the terms of second (and higher) order in $\delta$.
Namely, we calculate 
to the terms of order $\varepsilon \times \delta$
(i.e. the linear perturbation at the 1PN order).

Using a complex plane as the orbital one, 
we denote the relative position of the bodies as 
$\boldsymbol{r}_{I J} \to z_{I J} = r_{I J} e^{i \theta_{I J}}$.
The EIH equation of motion for $z_{12}$ becomes
\begin{align}
\frac{d^2 z_{12}}{d t^2} = F_{12} e^{i \theta_{12}} .
\end{align}
The left-hand side of this equation is
\begin{align}
\frac{d^2 z_{12}}{d t^2} &= \ell \omega_{\rm N}^2 
\left[ - \{ 1 + 2 \tilde{\omega}_{\rm PN} + \rho_{12} 
+ \delta (2 D \sigma_{12} + \chi_{12} - D^2 \chi_{12} 
+ 2 \tilde{\omega}_{\rm PN} D \sigma_{12} + 2 \rho_{12} D \sigma_{12} \right.
\notag\\
&~~~
\left.
+ 2 \tilde{\omega}_{\rm PN} \chi_{12}) \}
+ i \delta (2 D \chi_{12} + D^2 \sigma_{12} 
+ 2 \tilde{\omega}_{\rm PN} D \chi_{12} + \rho_{12} D^2 \sigma_{12}) 
\right] e^{i \theta_{12}} .
\end{align}
$F_{12}$ in the right-hand side of the equation of motion can be expanded as
\begin{align}
F_{12} = F_{{\rm N}\, 12} + \varepsilon F_{{\rm PN}\, 12} 
+ \delta F_{{\rm N per}\, 12} + \varepsilon \delta F_{{\rm PN per}\,12} ,
\end{align}
where $F_{{\rm N}\, 12}$ and $F_{{\rm PN}\, 12}$ are 
the unperturbed Newtonian and PN terms, respectively, 
and $F_{{\rm N per}\, 12}$ and $F_{{\rm PN per}\, 12}$ are 
the perturbed Newtonian and PN terms, respectively.
These are
\begin{align}
F_{{\rm N}\, 12} &= - \frac{M}{\ell^2} , 
\label{Appeq:fn12}
\\
F_{{\rm PN}\, 12} &= \frac{1}{24} \frac{M}{\ell^2}
(45 \nu_2^2  +  54 \nu_1 \nu_2 - 60 \nu_2  +  45 \nu_1^2 - 60 \nu_1  +  97) , 
\\
F_{{\rm N per}\, 12} &=
\frac12 \frac{M}{\ell^2} \left[
3 \nu_3 (\chi_{23}  +  \chi_{31}) + 2 ( 2  -  3 \nu_3) \chi_{12}
\right]
 + 
i \frac{3 \sqrt{3}}{2} \frac{M}{\ell^2} \nu_3 (\chi_{31} - \chi_{23}) , 
\end{align}

\begin{align}
F_{{\rm PN per}\, 12} &=
\frac{1}{16} \frac{M}{\ell^2} \biggl[
- 2 (54 \nu_2^3 + 108 \nu_1 \nu_2^2 - 86 \nu_2^2 + 108 \nu_1^2 \nu_2 
- 82 \nu_1 \nu_2 + 167 \nu_2 + 54 \nu_1^3 
\notag\\
&~~~
- 86 \nu_1^2 
+ 167 \nu_1 - 29) \chi_{12} 
- \nu_3
(45 \nu_2^2 + 108 \nu_1 \nu_2 + 8 \nu_2 + 90 \nu_1^2 - 108 \nu_1 + 150) 
\notag\\
&~~~
\times \chi_{23}
- \nu_3 
(90 \nu_2^2 + 108 \nu_1 \nu_2 - 108 \nu_2 + 45 \nu_1^2 + 8 \nu_1 + 150) \chi_{31}
+ 8 (\nu_2^3 - \nu_1 \nu_2^2 
\notag\\
&~~~
+ 2 \nu_2^2 - \nu_1^2 \nu_2 - 4 \nu_1 \nu_2 
- 7 \nu_2 + \nu_1^3 + 2 \nu_1^2 - 7 \nu_1) D \sigma_{12} 
- 2 \nu_3 (\nu_2^2 + 22 \nu_1 \nu_2 + 4 \nu_2 
\notag\\
&~~~
+ 4 \nu_1^2 + 2 \nu_1 + 8)
D \sigma_{23} 
- 2 \nu_3 
(4 \nu_2^2 + 22 \nu_1 \nu_2 + 2 \nu_2 + \nu_1^2 + 4 \nu_1 + 8) D \sigma_{31} 
\notag\\
&~~~
+ 8 \sqrt{3} \nu_3 (\nu_1 - \nu_2)(3 - \nu_3) D \chi_{12}
- 2 \sqrt{3} \nu_3 
(9 \nu_2^2 - 4 \nu_1 \nu_2 - 4 \nu_2 + 4 \nu_1^2 + 6 \nu_1 
\notag\\
&~~~
- 16) D \chi_{23}
+ 2 \sqrt{3} \nu_3 
(4 \nu_2^2 - 4 \nu_1 \nu_2 + 6 \nu_2 + 9 \nu_1^2 - 4 \nu_1 - 16) D \chi_{31}
\biggr]
\notag\\
&~~~
+ i \frac{1}{16} \frac{M}{\ell^2} 
\biggl[
- 12 \sqrt{3} \nu_3 (\nu_1 - \nu_2) (3 - \nu_3) \chi_{12} 
+ \sqrt{3} \nu_3 
(57 \nu_2^2 + 36 \nu_1 \nu_2 - 24 \nu_2 
\notag\\
&~~~
+ 42 \nu_1^2 - 12 \nu_1 + 130) \chi_{23} 
- \sqrt{3} \nu_3 
(42 \nu_2^2 + 36 \nu_1 \nu_2 - 12 \nu_2 + 57 \nu_1^2 - 24 \nu_1 
\notag\\
&~~~
+ 130) \chi_{31} 
- 8 \sqrt{3} \nu_3 (\nu_1 - \nu_2) (\nu_1 + \nu_2) D \sigma_{12} 
+ 2 \sqrt{3} \nu_3 
(\nu_2^2 - 12 \nu_1 \nu_2 + 14 \nu_2 
\notag\\
&~~~
- 4 \nu_1^2 + 10 \nu_1 + 8) D \sigma_{23} 
+ 2 \sqrt{3} \nu_3 
(4 \nu_2^2 + 12 \nu_1 \nu_2 - 10 \nu_2 - \nu_1^2 - 14 \nu_1 - 8) D \sigma_{31} 
\notag\\
&~~~
- 8 (3 \nu_2^3 + 9 \nu_1 \nu_2^2 - 6 \nu_2^2 + 9 \nu_1^2 \nu_2 - 8 \nu_1 \nu_2
- 5 \nu_2 + 3 \nu_1^3 - 6 \nu_1^2 - 5 \nu_1) D \chi_{12}
\notag\\
&~~~
- 2 \nu_3 (9 \nu_2^2 + 30 \nu_1 \nu_2 + 10 \nu_2 + 12 \nu_1^2 - 18 \nu_1 - 16)
 D \chi_{23}
- 2 \nu_3 (12 \nu_2^2 + 30 \nu_1 \nu_2 
\notag\\
&~~~
- 18 \nu_2 + 9 \nu_1^2 + 10 \nu_1 - 16) D \chi_{31} 
\biggr] .
\label{Appeq:fpnp12}
\end{align}
For exchanging indices between $1$ and $2$ such as 
$\nu_1 \leftrightarrow \nu_2$ and $\chi_{31} \leftrightarrow \chi_{23}$, 
we have the symmetry/antisymmetry in the real/imaginary parts of 
Eqs. \eqref{Appeq:fn12} - \eqref{Appeq:fpnp12}.
This (anti)symmetry may ensure the form of the equations,
though they are rather complicated.
We can obtain the EIH equations of motion for $z_{23}$ and $z_{31}$ 
by the cyclic manipulations as $1 \to 2 \to 3 \to 1$.
Since the unperturbed terms in the equations give 
the PN triangular equilibrium solution, 
we focus on the perturbed terms.
It is convenient to transform the variables as \cite{Routh}
\begin{align}
\chi_{23} &=
\frac12 \left[
( 1 - 3 \rho_{12} ) \chi_{31} 
+ ( 1 - 3 \rho_{31} ) \chi_{12} 
+ \sqrt{3} (1 - \rho_{23} ) \psi_{23} \right] , 
\label{ap:chi23} \\
X &\equiv \chi_{31} - \chi_{12} .
\label{ap:X}
\end{align}
Using the relations 
Eqs. \eqref{ap:s23}, \eqref{ap:s31}, \eqref{ap:chi23}, and \eqref{ap:X},
we obtain the equations of motion 
\eqref{C6eq:z12Re} - \eqref{C6eq:z31Im} for perturbations.

\clearpage

\begin{thebibliography}{99}
\bibitem{Goldstein}
H. Goldstein, {\it Classical Mechanics} 
(Addison-Wesley, MA, 1980). 
\bibitem{Danby}
J. M. A. Danby, {\it Fundamentals of Celestial Mechanics} 
(William-Bell, VA, 1988). 
\bibitem{Marchal}
C. Marchal, {\it The Three-Body Problem} 
(Elsevier, Amsterdam, 1990). 
\bibitem{CWV}
M. Connors, P. Wiegert and C. Veillet, Nature {\bf 475}, 481 (2011).
\bibitem{Krefetz}
E. Krefetz, Astron. J {\bf 72}, 471 (1967).
\bibitem{Maindl}
T. I. Maindl, 
{\it Completing the Inventory of the Solar System, Astronomical Society of
the Pacific Conference Proceedings}, edited by T.W. Rettig and
J.M. Hahn, {\bf 107}, 147 (1996).
\bibitem{Asada}
H. Asada, Phys. Rev. D {\bf 80} 064021 (2009). 
\bibitem{SM}
N. Seto and T. Muto, Phys. Rev. D {\bf 81} 103004 (2010).
\bibitem{Schnittman}
J. D. Schnittman, Astrophys. J. {\bf 724} 39 (2010). 
\bibitem{THA}
Y. Torigoe, K. Hattori, and H. Asada, 
Phys.\ Rev.\ Lett.\  {\bf 102}, 251101 (2009).
\bibitem{Nordtvedt}
K. Nordtvedt,  
Phys. Rev. {\bf 169} 1014 (1968).
\bibitem{BLS}
O. Blaes, M. H. Lee, and A. Socrates, Astrophys. J. {\bf 578}, 775 (2002).
\bibitem{MH}
M. C. Miller and D. P. Hamilton, Astrophys. J. {\bf 576}, 894 (2002).
\bibitem{Wen}
L. Wen, Astrophys. J. {\bf 598}, 419 (2003)
\bibitem{Thompson}
T. A. Thompson, Astrophys. J. {\bf 741}, 82 (2011).
\bibitem{Seto}
N. Seto, Phys. Rev. Lett. {\bf 111}, 061106 (2013).
\bibitem{Ransom}
S. M. Ransom {\it et al.}, Nature {\bf 505}, 520 (2014).
\bibitem{YA1}
K. Yamada and H. Asada, Phys. Rev. D {\bf 82}, 104019 (2010).
\bibitem{YA2}
K. Yamada and H. Asada, Phys. Rev. D {\bf 83}, 024040 (2011).
\bibitem{IYA}
T. Ichita, K. Yamada, and H. Asada, Phys. Rev. D {\bf 83}, 084026 (2011).
\bibitem{YA3}
K. Yamada and H. Asada, Phys. Rev. D {\bf 86}, 124029 (2012).
\bibitem{Gascheau}
G. Gascheau, C. R. Acad. Sci. {\bf 16}, 393 (1843).
\bibitem{Routh}
R. J. Routh, Proc. Lond. Math. Soc. {\bf 6}, 86 (1875).
\bibitem{DP}
C. N. Douskos and E.A. Perdios,
Celest. Mech. Dyn. Astron. {\bf 82}, 317 (2002).
\bibitem{SB1}
J. Singh and N. Bello, Astrophys. Space Sci., {\bf 351}, 483, (2014).
\bibitem{SB2}
J. Singh and N. Bello, Astrophys. Space Sci., {\bf 351}, 491, (2014).
\bibitem{MTW}
C. W. Misner, K. S. Thorne, and J. A. Wheeler, 
{\it Gravitation}, 
(Freeman, New York, 1973).
\bibitem{LL}
L. D. Landau and E. M. Lifshitz, 
{\it The Classical Theory of Fields} 
(Oxford, Pergamon 1962).
\bibitem{Will}
C. M. Will, 
{\it Theory and experiment in gravitational physics} 
(Cambridge University, New York, 1993).
\bibitem{AFH}
H. Asada, T. Futamase, and P. Hogan, 
{\it Equations of Motion in General Relativity} 
(Oxford University, New York, 2011).



\end{thebibliography}
\end{document}